\documentclass[aps,pra,floatfix,twocolumn,superscriptaddress,twoside]{revtex4-1}
\usepackage{amsmath,amssymb,amsfonts}
\usepackage{graphicx}
\usepackage[english]{babel}
\usepackage{psfrag}
\usepackage{color}
\usepackage{tabularx}
\usepackage{times}
\usepackage{hyperref}
\usepackage{float}

\begin{document}

\title{Extreme Suppression of Antiferromagnetic Order and Critical Scaling \\ in a Two-Dimensional Random Quantum Magnet}

\author{Wenshan Hong}
\affiliation{Beijing National Laboratory for Condensed Matter Physics, Institute of Physics, Chinese Academy of Sciences, Beijing 100190, China}
\affiliation{School of Physical Sciences, University of Chinese Academy of Sciences, Beijing 100190, China}

\author{Lu Liu}
\affiliation{Beijing National Laboratory for Condensed Matter Physics, Institute of Physics, Chinese Academy of Sciences, Beijing 100190, China}

\author{Chang Liu}
\affiliation{Beijing National Laboratory for Condensed Matter Physics, Institute of Physics, Chinese Academy of Sciences, Beijing 100190, China}
\affiliation{School of Physical Sciences, University of Chinese Academy of Sciences, Beijing 100190, China}

\author{Xiaoyan Ma}
\affiliation{Beijing National Laboratory for Condensed Matter Physics, Institute of Physics, Chinese Academy of Sciences, Beijing 100190, China}
\affiliation{School of Physical Sciences, University of Chinese Academy of Sciences, Beijing 100190, China}

\author{Akihiro Koda}
\affiliation{Institute of Materials Structure Science, High Energy Accelerator Research Organization (KEK-IMSS),1-1 Oho, Tsukuba 305-0801, Japan}
\affiliation{Department of Materials Structure Science, Sokendai (The Graduate University for Advanced Studies), Tsukuba, Ibaraki, 305-0801, Japan}

\author{Xin Li}
\affiliation{Key Laboratory of Neutron Physics and Institute of Nuclear Physics and Chemistry, China Academy of Engineering Physics, Mianyang 621999, China}

\author{Jianming Song}
\affiliation{Key Laboratory of Neutron Physics and Institute of Nuclear Physics and Chemistry, China Academy of Engineering Physics, Mianyang 621999, China}

\author{Wenyun Yang}
\affiliation{State Key Laboratory for Mesoscopic Physics, School of Physics, Peking University, Beijing, 100871, China}

\author{Jinbo Yang}
\affiliation{State Key Laboratory for Mesoscopic Physics, School of Physics, Peking University, Beijing, 100871, China}

\author{Peng Cheng}
\affiliation{Department of Physics and Beijing Key Laboratory of Opto-electronic Functional Materials \& Micro-nano Devices, Renmin University of China, Beijing 1\
00872, China}

\author{Hongxia Zhang}
\affiliation{Department of Physics and Beijing Key Laboratory of Opto-electronic Functional Materials \& Micro-nano Devices, Renmin University of China, Beijing 1\
00872, China}

\author{Wei Bao}
\affiliation{Department of Physics and Beijing Key Laboratory of Opto-electronic Functional Materials \& Micro-nano Devices, Renmin University of China, Beijing 1\
00872, China}
\affiliation{Department of Physics, City Univesity of Hong Kong, Kowloon, Hong Kong}

\author{Xiaobai Ma}
\affiliation{Department of Nuclear Physics, China Institute of Atomic Energy, Beijing, 102413, China}

\author{Dongfeng Chen}
\affiliation{Department of Nuclear Physics, China Institute of Atomic Energy, Beijing, 102413, China}

\author{Kai Sun}
\affiliation{Department of Nuclear Physics, China Institute of Atomic Energy, Beijing, 102413, China}

\author{Wenan Guo}
\affiliation{Department of Physics, Beijing Normal University, Beijing 100875, China}
\affiliation{Beijing Computational Science Research Center, Beijing 100193, China}

\author{Huiqian Luo}
\affiliation{Beijing National Laboratory for Condensed Matter Physics, Institute of Physics, Chinese Academy of Sciences, Beijing 100190, China}
\affiliation{Songshan Lake Materials Laboratory, Dongguan, Guangdong 523808, China}

\author{Anders W. Sandvik}
\email{sandvik@bu.edu}
\affiliation{Department of Physics, Boston University, 590 Commonwealth Avenue, Boston, Massachusetts 02215, USA}
\affiliation{Beijing National Laboratory for Condensed Matter Physics, Institute of Physics, Chinese Academy of Sciences, Beijing 100190, China}

\author{Shiliang Li}
\email{slli@iphy.ac.cn}
\affiliation{Beijing National Laboratory for Condensed Matter Physics, Institute of Physics, Chinese Academy of Sciences, Beijing 100190, China}
\affiliation{School of Physical Sciences, University of Chinese Academy of Sciences, Beijing 100190, China}
\affiliation{Songshan Lake Materials Laboratory, Dongguan, Guangdong 523808, China}

\date{July 24, 2020}

\begin{abstract}
Sr$_2$CuTeO$_6$ is a square-lattice N\'eel antiferromagnet with superexchange between first-neighbor $S=1/2$ Cu spins mediated by plaquette centered Te ions.
Substituting Te by W, the affected impurity plaquettes have predominantly second-neighbor interactions, thus causing local magnetic frustration. Here we report
a study of Sr$_2$CuTe$_{1-x}$W$_x$O$_6$ using neutron diffraction and $\mu$SR techniques, showing that the N\'eel order vanishes already at $x = 0.025 \pm 0.005$.
We explain this extreme order suppression using a two-dimensional Heisenberg spin model, demonstrating that a W-type impurity induces a deformation of the order
parameter that decays with distance as $1/r^2$ at temperature $T=0$. The associated logarithmic singularity leads to loss of order for any $x>0$. Order for small
$x>0$ and $T>0$ is induced by weak interplane couplings. In the nonmagnetic phase of Sr$_2$CuTe$_{1-x}$W$_x$O$_6$, the $\mu$SR relaxation rate exhibits
quantum critical scaling with a large dynamic exponent, $z \approx 3$, consistent with a random-singlet state.
\end{abstract}

\maketitle

A central theme in modern condensed matter physics is the evolution of two-dimensional (2D) quantum antiferromagnets upon doping, as epithomized by the
high-T$_c$ cuprates with charge carriers introduced into the CuO$_2$ layers through off-layer doping \cite{Lee06,Chatterjee17}. In-plane static impurities
have also been studied, e.g., non-magnetic Zn substituting the spin $S=1/2$ carrying Cu ions \cite{Vajk02,Sandvik02,Liu09}. In general, impurities and
random frustrated couplings in a quantum magnet will eventually destroy any order and may induce not yet fully understood disordered states, e.g., quantum spin 
glasses \cite{Ye93,Oitmaa01,Dey20}, spin fluids \cite{Sachdev93}, valence-bond glasses \cite{Tarzia08,Singh10}, and random-singlet (RS) states
\cite{Bhatt82,Fisher94,Motrunich00,Lin03,Laumann12,Watanabe14,UematsuK18,Kimchi18a,Kimchi18b,Kawamura19,Liu18,Liu20,Ren20}.

\begin{figure}[b]
\includegraphics[width=83mm]{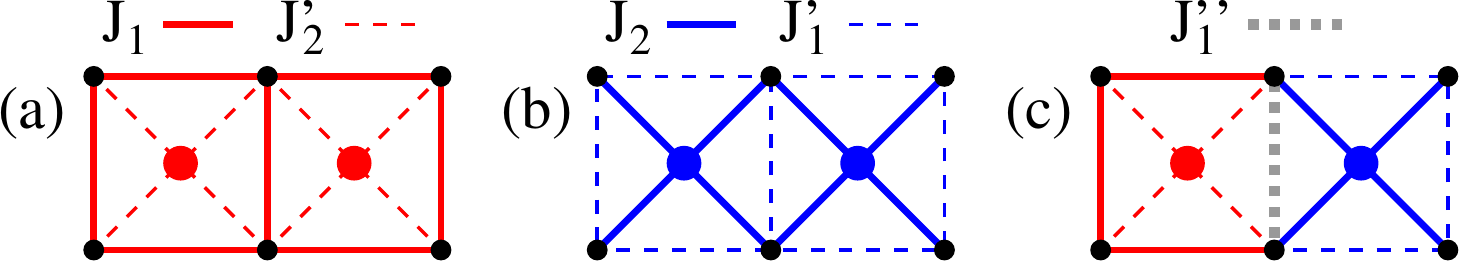}
\caption{2D Heisenberg couplings $J_{ij}{\bf S}_i \cdot {\bf S}_j$ in  Sr$_2$CuTe$_{1-x}$W$_x$O$_6$. The
small black circles represent the $S=1/2$ carrying Cu ions, while red and blue circles correspond to Te and W ions, respectively. The dominant couplings
mediated by Te in (a) and W in (b) are first-neighbor $J_1$ (solid red lines) and second-neighbor $J_2$ (solid blue lines), with $J_1 \approx J_2 \approx 8$
meV \cite{MustonenO18a,Katukuri20}. The couplings $J'_1$ and $J'_2$ indicated by the thin dashed lines are roughly 10\% of the dominant couplings. The
first-neighbor coupling $J''_1$ on links between Te and W ions, the gray dashed line in (c), is about $4\%$ of $J_1$ \cite{Katukuri20}.}
\label{fig:j1j2}
\end{figure}

We here report $\mu$SR and neutron diffraction experiments on Sr$_2$CuTe$_{1-x}$W$_x$O$_6$, which at $x=0$
realizes the 2D $S=1/2$ antiferromagnetic (AFM) Heisenberg model with predominantly first-neighbor interactions $J_1$ generated through
superexchange via Te ions at the centers of the plaquettes of $2\times 2$ Cu ions \cite{KogaT16,BabkevichP16}; see Fig.~\ref{fig:j1j2}(a).
At $x=1$, the W ions instead mediate second-neighbor superexchange in the affected plaquettes, Fig.~\ref{fig:j1j2}(b), with
$J_2 \approx J_1$ \cite{VasalaS14,VasalaS14b,WalkerHC16}. An intriguing magnetically disordered state exists within a window
$[x_{{\rm c}1},x_{{\rm c}2}]$, with $x_{{\rm c}1} \approx 0.1$ and $x_{{\rm c}2} \approx 0.6$ estimated \cite{MustonenO18a,WatanabeM18,MustonenO18b}.
The ability to tune the disorder and frustration by $x$ offers unique opportunities to systematically study frustrated plaquette impurities of the
$J_2$ type illustrated in Fig.~\ref{fig:j1j2}(c) for small $x$ and the subsequent randomness-induced non-magnetic state for larger $x$.

We here demonstrate destruction of the N\'eel order in Sr$_2$CuTe$_{1-x}$W$_x$O$_6$ at $x_{{\rm c}1} = 0.025 \pm 0.005$, far below the previous estimate.
We explain this dramatic order suppression using a classical Heisenberg model with random W and Te ions. Here 2D N\'eel order at temperature $T = 0$ is
destroyed even at infinitesimal $x$, due to a logarithmic singularity caused by the single-impurity deformation of the spin texture. Order at $x>0$ and $T>0$
is stabilized by weak inter-layer couplings. The columnar AFM state extending from $x= 1$ is much more robust, which also can be explained by the classical
model. In the non-magnetic phase, the neutron diffraction measurements reveal short-range N\'eel correlations and the $\mu$SR relaxation rate exhibits
quantum-critical scaling with dynamic exponent $z>2$, both consistent with recent predictions for the 2D RS state \cite{Liu18,Liu20}.

{\it Experiments.}---Polycrystalline Sr$_2$CuTe$_{1-x}$W$_x$O$_6$ samples were synthesized as described previously
\cite{KogaT16,BabkevichP16,VasalaS14,WalkerHC16}. The experiments were carried out at J-PARC ($\mu$SR) and China Advanced Research Reactor and
Key Laboratory of Neutron Physics and Institute of Nuclear Physics and Chemistry, China (neutron diffraction); see also Supplemental Material \cite{sm}.

\begin{figure}[t]
\includegraphics[width=84mm]{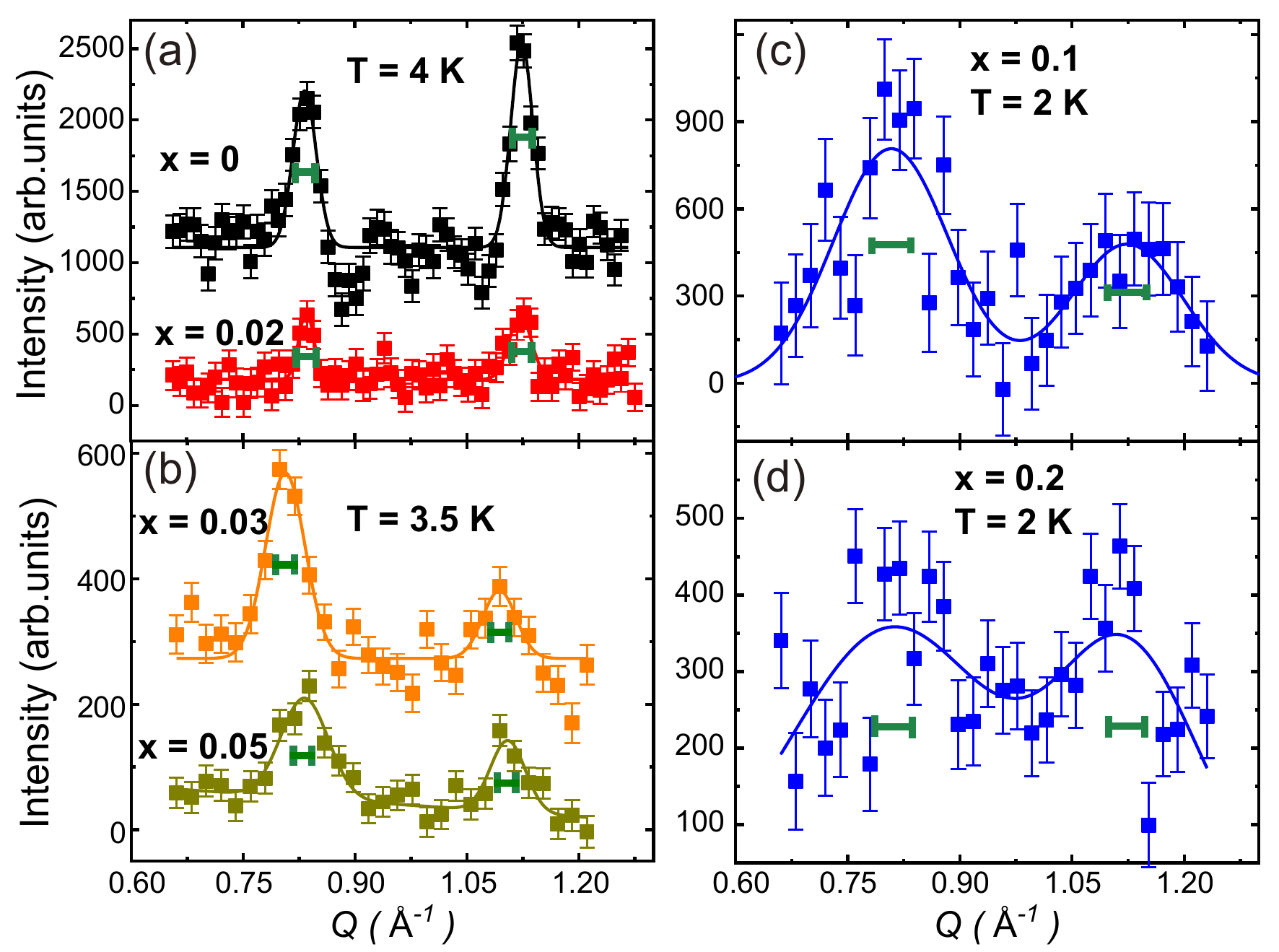}
\caption{Neutron diffraction results for (a) $x$ = 0 and 0.02, (b) 0.03 and 0.05 (c) 0.1, and (d) 0.2. The peaks correspond to wave-vectors $q$ = (1/2,1/2,0) 
and (1/2,1/2,1) in the tetragonal magnetic Brillouin zone, indicating dominant N\'eel AFM order ($x$ = 0 and 0.02) and short-range correlations ($x \geq$ 0.03). 
Data at $T$ = 40 K have been subtracted as background. The $x$ = 0 and 0.03 values have been shifted vertically for clearity. The curves are Gaussian fits
and the green bars indicate the instrumental resolutions.}
\label{fig:neutron}
\end{figure}

Figure \ref{fig:neutron}  shows our neutron diffraction results. Resolution limited magnetic peaks are observed at $x$ = 0 in Fig.~\ref{fig:neutron}(a),
consistent with N\'eel AFM order \cite{KogaT16,MustonenO18a}. We have also confirmed (Supplemental Material \cite{sm}) columnar AFM order
\cite{WatanabeM18,MustonenO18b} for $x \in$ [0.7, 1]. The W doped sample with $x$ = 0.02, Fig.~\ref{fig:neutron}(a), is still ordered,
with resolution limited peaks (corresponding to a correlation length $>$ 180 {\AA} $\approx$ 35 lattice spacings). The broader peaks for $x \geq$ 0.03
in Figs.~\ref{fig:neutron}(b)-\ref{fig:neutron}(d) indicate the loss of long-range order between $x$ = 0.02 and 0.03. At $x$ = 0.1 the correlation
length is still about 40 {\AA}.

\begin{figure}[t]
\includegraphics[width=85mm]{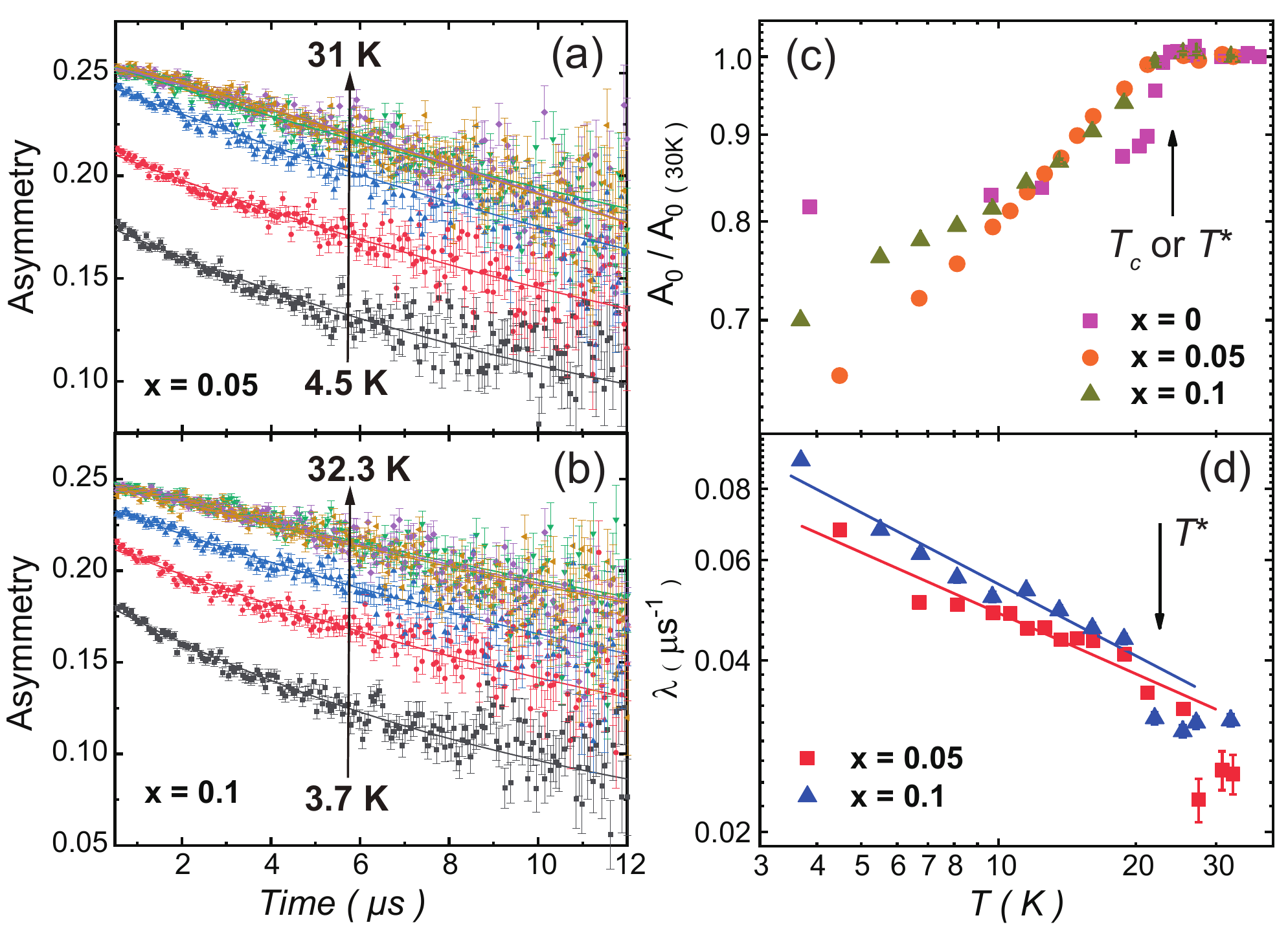}
\caption{Time-dependent zero-field $\mu$SR spectra for (a) $x$ = 0.05 and (b) $x=$ 0.1 samples at different temperatures (the highest and lowest indicated)
along with fits to Eq.~(\ref{atform}). (c) Temperature dependent $\mu$SR asymmetry for $x$ = 0, 0.05, and 0.1, normalized by the values at $T$ = 30 K. 
(d) Temperature dependent relaxation rate $\lambda$ for $x$ = 0.05 and 0.1. The fitted lines correspond to critical scaling, $\lambda \approx T^{-\gamma}$,
with $\gamma$ = 0.35 $\pm$ 0.03 ($x$ = 0.05) and 0.42 $\pm$ 0.03 ($x$ = 0.1).}
\label{fig:usr}
\end{figure}

The $\mu$SR asymmetry $A(t)$ was fitted to
\begin{equation}
A(t) = A_0 {\rm exp}(-\lambda t)G_z(t) + A_{\rm BG},
\label{atform}
\end{equation}
\noindent where $A_0$ is the initial asymmetry, $\lambda$ the relaxation rate of the muon spins, $A_{\rm BG}$ the constant background, and $G_z(t)$ 
the Kubo-Toyabe function \cite{HayanoRS79}. The function $A(t)$ cannot actually describe the complete muon spectra of the magnetically ordered samples. It
has already been shown that, for columnar AFM ordered systems at $x$ = 1, 0.9, and 0.8, the asymmetry initially drops very rapidly and oscillates
\cite{VasalaS14b,MustonenO18b}. These features take place within 1 $\mu s$, beyond the resolution of our measurements. Instead, Eq.~(\ref{atform})
describes the relaxation at longer times and $A_0$ is close to the asymmetry after the rapid initial drop. While the fits of Eq.~(\ref{atform})
are not perfect for the long-range ordered samples (Supplemental Material \cite{sm}), the form describes the data for $x$ = 0.05 and 0.1
very well, as shown in Figs.~\ref{fig:usr}(a) and \ref{fig:usr}(b).

\begin{figure}[t]
\includegraphics[width=68mm]{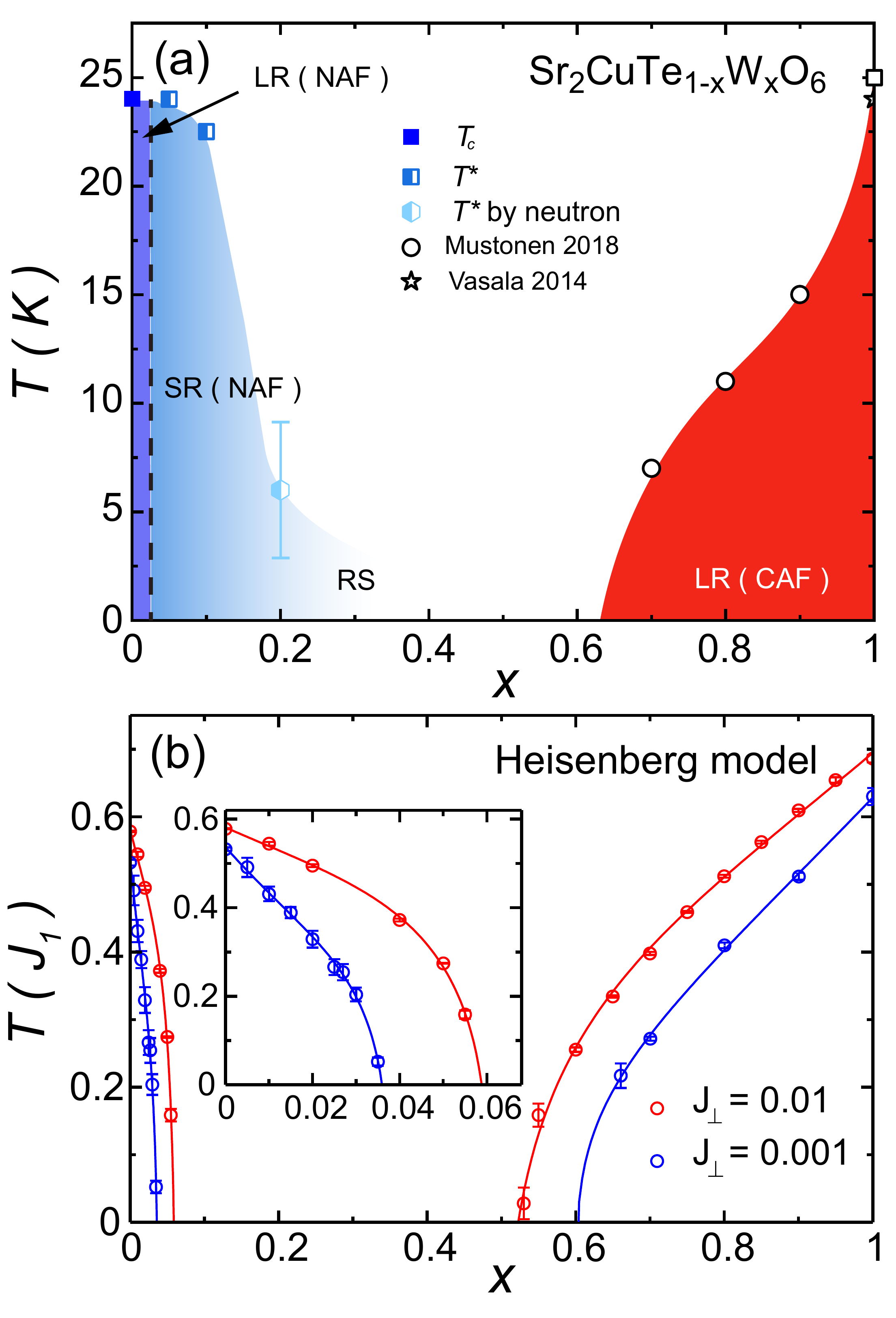}
\caption{(a) Magnetic phase diagram of Sr$_2$CuTe$_{1-x}$W$_x$O$_6$. NAF and CAF denote N\'eel and columnar AFM correlations, respectively,
either short-range (SR) or long-range (LR). The ordering temperature $T_{\rm c}$ and characteristic short-range correlation temperature $T^*$
were determined by $\mu$SR measurements, except for $T^*$ of the $x = 0.2$ sample, which was obtained (Supplemental Material \cite{sm}) by neutron
diffraction (b) Transition temperatures of the classical Heisenberg model of coupled layers, determined using Monte Carlo simulations. In the
notation of Fig.~\ref{fig:j1j2} the 2D couplings are $J_1=J_2=1$, $J'_1=J'_2=0.1$, and $J''_1=0$. Two different interlayer couplings are
used; $J_\perp=10^{-2}$ and $10^{-3}$. Curves are drawn through the data points as guides to the eye.}
\label{fig:phases}
\end{figure}

The temperature dependent $A_0$ is graphed in Fig.~\ref{fig:usr}(c) for $x$ = 0, 0.05 and 0.1. A sharp change is observed at the previously known ordering
temperature $T_{\rm c}$ at $x$ = 0 \cite{KogaT16,BabkevichP16}. In contrast, in the $x$ = 0.05 and 0.1 samples $A_0$ only decreases slowly below a characteristic
temperature $T^*$. This behavior reflects gradual changes of the local fields as a result of the onset of short-range magnetic correlations but no ordering,
which is consistent with the neutron results in Figs. \ref{fig:neutron}(b) and \ref{fig:neutron}(c). It should be noted that the value of $A_0$ for $x$ = 0 at
low temperatures is about 4/5 of that above $T_{\rm c}$, while in the case of $x$ = 1 it is only 1/3 \cite{VasalaS14,sm}. It is beyond the scope
of this work to explain the detailed form of $A_0$; some additional analysis is provided in Supplementary Material \cite{sm}.
	
Fig.~\ref{fig:usr}(d) shows the temperature dependence of the relaxation rate $\lambda$ for $x$ = 0.05 and 0.1. Power-law behaviors reflect quantum-critical
scaling in what is likely the RS phase. As explained in Supplemental Material, standard scaling arguments \cite{Fisher89,Chubukov93} in combination with a
constraint imposed by the recently discovered $1/r^2$ form of the spin correlations in the 2D RS phase \cite{Liu18,Liu20,Ren20} can be used to derive the
form $\lambda \propto T^{-\gamma}$ with $\gamma = 1-2/z$, where $z$ is the dynamic exponent. The values of $\gamma$ extracted from the fits in Fig.~\ref{fig:usr}(d)
correspond to $z$ = 3.0 $\pm$ 0.2 for $x$ = 0.05 and $z$ = 3.5 $\pm$ 0.3 for $x$ = 0.1. These values conform with the expectations in the RS phase,
where $z$ equals $2$ at the N\'eel--RS transition and grows upon moving into the RS phase \cite{Liu18}. It should be noted that the value of $A_{\rm BG}$
in Eq.~(\ref{atform}) somewhat affects the determination of $\gamma$ but we consistently find power law behavior of $\lambda$ and $z(x=0.1) > z(x=0.05)$
(further discussed in Supplemental Material \cite{sm}). We note that the low-temperature $\mu$SR relaxation in quasi-2D spin glasses is very
different \cite{Yadav19}.

Combining our $\mu$SR and neutron results with previous works, the magnetic phase diagram of Sr$_2$CuTe$_{1-x}$W$_x$O$_6$ is shown in Fig.~\ref{fig:phases}(a).
The columnar order at $x$ = 1 is robust even for large Te substitution, which is indicative of only minor effects of magnetic frustration and remaining large connected
ordered regions. The mean order parameter may then be gradually reduced in a way similar to diluted systems \cite{CollinsMF89}. In contrast, introducing W in the 
$x = 0$ sample rapidly destroys the N\'eel order at $x_{{\rm c}1} =0.025 \pm 0.005$. Short-range correlations with N\'eel structure still remain at low 
temperatures even at $x=0.2$ based on our neutron-diffraction experiments and likely persist throughout what we argue is the 2D RS phase. 

{\it Modeling.}---The width of the N\'eel phase in Fig.~\ref{fig:phases}(a) is less than $1/3$ of the previous estimates \cite{MustonenO18a,WatanabeM18,MustonenO18b}.
The N\'eel phase at finite W doping being narrower than the columnar phase at finite Te doping can be understood already at the classical level
with the dominant Heisenberg coupling constants $J_1$ and $J_2$ in Fig.~\ref{fig:j1j2}: Introducing a single Te impurity in the $J_2$-coupled columnar
system, we simply lose the $J_2$ couplings in the affected plaquette and there is only weak frustration from the much smaller $J'_1$ and $J''_1$ couplings. However,
with a W impurity in the $J_1$-dominated N\'eel state the two new $J_2$ bonds are completely frustrated. To quantitatively understand the extremely narrow N\'eel
phase requires further insights.

Ideally, we would like to carry out calculations with the full quantum mechanical Heisenberg Hamiltonian. Even though progress has been made on some
frustrated 2D quantum magnets with density-matrix renormalization group (DMRG) \cite{Verresen18} and tensor-product \cite{Chen19} methods, including
Heisenberg systems with random couplings \cite{Ren20}, in practice calculations for frustrated systems are still challenging and it would be hard
to extract a reliable phase diagram. However, we have found that already the classical Heisenberg model can explain the extreme fragility of the N\'eel
state to W-plaquette impurities and also gives an overall reasonable phase diagram.

The long-range N\'eel order at $T=0$ in the 2D Heisenberg model with uniform exchange $J_1 {\bf S}_i \cdot {\bf S}_j$ on all first neighbors $(i,j)$ is
destroyed by thermal fluctuations at $T>0$ \cite{Mermin66,Chakravarty89}. In weakly coupled planes of classical or quantum spins, $T_c \propto J_1\ln^{-1}(J_1/J_\perp)$,
where $J_\perp$ is the coupling between spins in adjacent planes \cite{Irkhin98,Sengupta03}. Since a quantum magnet with AFM order or a long correlation
length behaves in many respects as a ``renormalized classical'' system \cite{Chakravarty89}, the initial effects of doping the $x=0$ and $x=1$ system should
be captured correctly by a classical model, up to $O(1)$ factors.

In the notation of Fig.~\ref{fig:j1j2}, we set the 2D couplings to $J_1=J_2=1$, $J'_1=J'_2=0.1$, and $J''_1=0$, with $|\mathbf{S}_i|=1$. For coupled
planes we consider $J_\perp=10^{-2}$ and $10^{-3}$. We used standard Monte Carlo methods for frustrated Heisenberg models \cite{Alonso96,Lee07}, with Binder
cumulant techniques \cite{Sandvik10} for extracting $T_c$ at fixed $x$, based on averages over several hundred realizations of the random W and Te
plaquettes on systems with up to $72\times 72\times 18$ spins. The resulting infinite-size extrapolated phase boundaries are shown in
Fig.~\ref{fig:phases}(b). When comparing with the experiments, it should be noted that $T=25$ K corresponds roughly to $0.3$ in units of $J_1$ and that
$T_c$ in uniform coupled $S=1/2$ planes with $J_\perp$ of order $10^{-2}$ is lower by about $50\%$ than our classical result at $x=0$ \cite{Sengupta03}.
We expect quantum fluctuations to shrink the ordered phases also in the $x$ direction, and the differences between the numerical and experimental
results for the columnar phase boundary should also be due to quantum effects (and possibly weak interactions beyond those included here).

As seen in Fig.~\ref{fig:phases}(b), upon changing $J_\perp$ from $10^{-2}$ to $10^{-3}$, $T_c$ at $x=0$ is only slightly reduced, as expected on account
of the logarithmic form discussed above. For $x>0$ the phase boundary drops more rapidly to zero for the smaller $J_\perp$, and
the size of the N\'eel phase is substantially smaller. A very narrow N\'eel phase with high sensitivity of the $T=0$ transition point to $J_\perp$ is not
expected within a simple picture of conventional local impurity suppression of the order \cite{CollinsMF89}. We therefore investigate the deformation
of the N\'eel order around a single impurity plaquette at $T=0$, which we have done by minimizing the energy with a combination of simulated annealing
and energy conserving spin moves.

\begin{figure}[t]
\includegraphics[width=75mm]{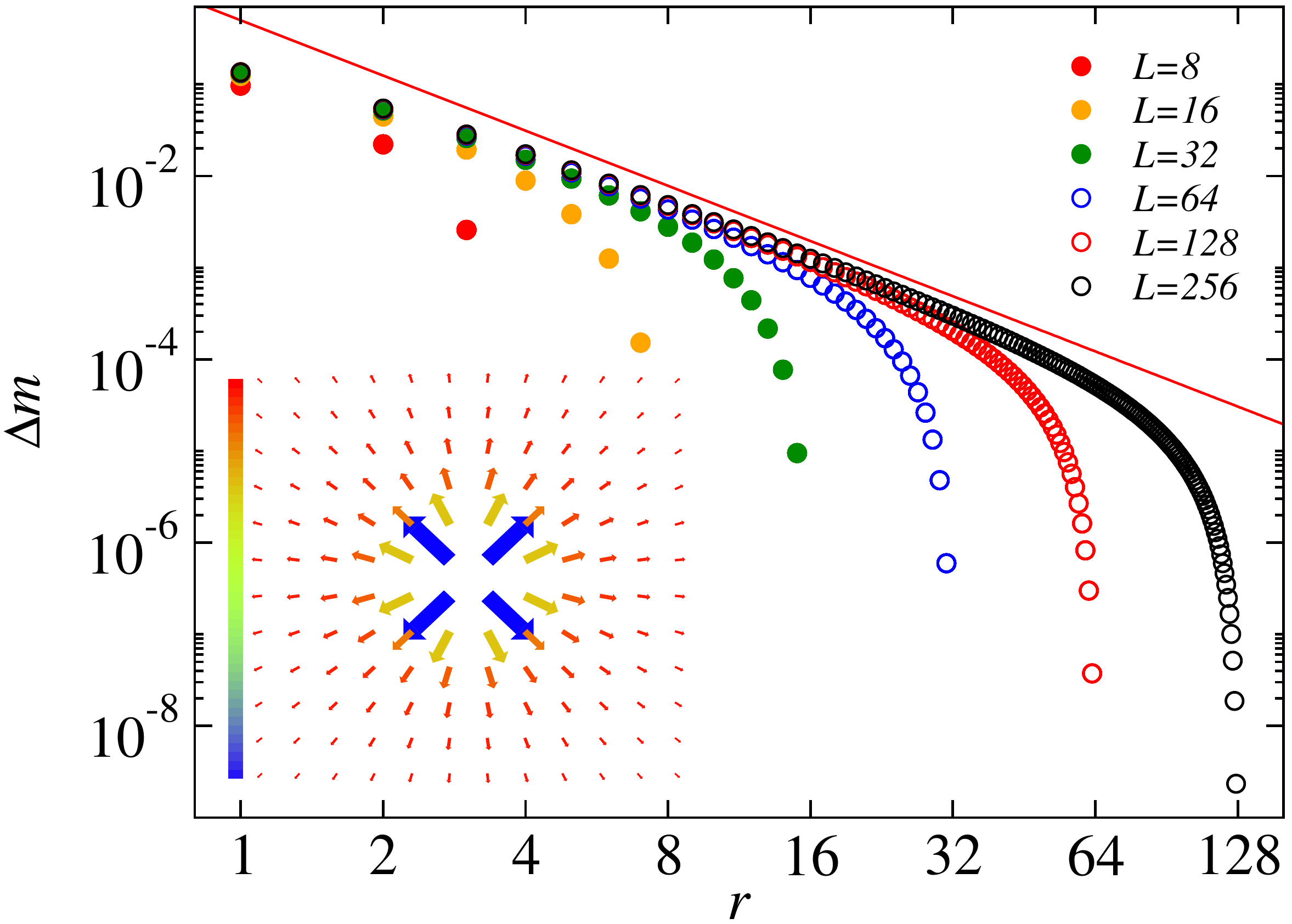}
\caption{Deformation of the order parameter of the classical Heisenberg model with a W-type plaquette impurity as defined in Fig.~\ref{fig:j1j2},
with the same couplings as in Fig.~\ref{fig:phases}. The deviation $\Delta m = 1-|S^z_i|$, where the $z$ direction is that of the bulk N\'eel order,
is shown vs the distance $r$ from the impurity along the $(1,0)$ lattice direction for several system sizes. The line shows the form $1/r^2$.
The inset shows the projection of the spins to the $xy$ spin plane, with the color coding corresponding to $m \in [0.49,1]$. The magnitude of
the $xy$ component decays as $1/r$ from its maximal value $\approx 0.87$ closest to the impurity. The behaviors correspond to an angular distortion
$\propto 1/r$.}
 \label{fig:deltam}
\end{figure}
 
The deviation $\Delta m$ of the local ordered moment from the bulk value is graphed in Fig.~\ref{fig:deltam} versus the distance $r$ from the impurity.
The form $\Delta m \propto 1/r^2$ causes a logarithmic divergence when integrated over $r$ (but the total energy cost of the deformation stays
constant, with the energy density decaying as $1/r^4$). This single-impurity response suggests that any impurity fraction $x>0$ destroys the long-range
order, and this is demonstrated explicitly in the Supplemental Material \cite{sm}. A similar fragility of non-colinear bulk order in the presence of
certain impurities was previously pointed out \cite{Dey20}, but the profound impact of the plaquette impurity (which can be understood as a composite
of two dipoles; see Supplemental Material \cite{sm}) on the colinear N\'eel state had not been anticipated.

For the weakly coupled planes in Fig.~\ref{fig:phases}(b), the N\'eel order is stabilized for a range of $x>0$ depending on $J_\perp/J_1$, but we have not
studied the functional form of $x_{c1}$ versus $J_\perp$. The disorder should be irrelevant at the $T>0$ phase transitions according to the Harris 
criterion \cite{Harris74,Chayes86}, and we expect standard three-dimensional O(3) universality. We do not have sufficient data for large systems to 
test the critical exponents. In an $S=1/2$ system such as Sr$_2$CuTe$_{1-x}$W$_x$O$_6$, quantum fluctuations should further suppress
the order and reduce $x_{{\rm c}1}$, and we expect the same type of logarithmic singularity as in the classical case when $J_\perp/J_1 \to 0$, on account of the
renormalized classical picture of the quantum N\'eel state \cite{Chakravarty89}.

{\it Discussion.}---The extreme effect of the W impurities in the N\'eel state was not captured by the density functional calculations
in Ref.~\cite{MustonenO18b}, which suggested destabilization of the N\'eel order for $x \approx$ 0.1-0.2 in Sr$_2$CuTe$_{1-x}$W$_x$O$_6$, significantly
above $x_{{\rm c}1} \approx 0.025$ found in our experiments. The mechanism we have uncovered here relies on a singular effect of frustrated plaquette
impurities in 2D, with weak 3D couplings pushing the transition from $x=0$ to to small $x>0$.

Once the N\'eel order vanishes, from the classical perspective a spin glass phase is expected \cite{Xu18,Dey20}. In the presence of strong quantum
fluctuations in $S=1/2$ systems, there is mounting evidence from model studies that the spin glass can be supplanted by an RS state
\cite{Kimchi18a,Liu18,Liu20,Ren20,Dey20}. A particular realization of the RS state amenable to large-scale quantum Monte Carlo calculations exhibits
criticality with a dynamic exponent $z \ge 2$ and dominant N\'eel-type spin correlations decaying with distance as $1/r^2$ at $T=0$ \cite{Liu18,Liu20}.
This form of the correlations was recently confirmed in a frustrated random-bond system with DMRG calculations \cite{Ren20}, thus further supporting
universal RS behavior. The significant staggered correlations well past the N\'eel phase in Sr$_2$CuTe$_{1-x}$W$_x$O$_6$, as revealed by our neutron difraction
experiments at $x=0.1$ and $0.2$, are thus expected within the RS scenario. Previous results at $x=0.5$ also showed remnants of N\'eel correlations
\cite{Katukuri20}. We here further demonstrated quantum-critical scaling of the $\mu$SR relaxation rate with varying $z>2$, as recently predicted in 
the 2D RS state \cite{Liu18,Liu20}.

It would be interesting to further test the proposed RS scaling forms experimentally in Sr$_2$CuTe$_{1-x}$W$_x$O$_6$. A re-analysis \cite{Liu18} of
susceptibility data for $x \ge 0.2$ \cite{WatanabeM18} supported the predicted form $\chi \propto T^{-\gamma}$ with $\gamma < 1$. Detailed inelastic
netron scattering studies would be very useful, but our attemps to grow large single-crystals have so far not been successful. With polycrystalline samples,
NMR experiments may be able to further elucidate the nature of the RS state and the N\'eel--RS transition. RS signatures were previously reported in
YbMgGaO$_4$ \cite{Kimchi18b} and $\alpha$-Ru$_{1-x}$Ir$_x$Cl$_3$ \cite{baek20}, but in addition to random frustration these materials have Dzyaloshinskii-Moriya
interactions and spin vacancies, respectively. Beyond its intrinsic importance, the 2D RS state should also be a useful benchmark for experiments on
potential uniform spin liquids \cite{Savary17,ZhouY17}, where it is often difficult \cite{Singh10,Li15,Ma18,Kimchi18b} to distinguish
between impurity physics and theoretically predicted properties of clean systems.

\begin{acknowledgments}
{\it Acknowledgments.}---W.H. and L.L contributed equally to this work. We would like to thank Oleg Sushkov for valuable comments. The research at Chinese institutions
is supported by the National Key R\&D Program of China (Grants No.~2017YFA0302900, No.~2016YFA0300502, No.~2018YFA0704201, No.~2016YFA0300604,
No.~2017YFA0303100), the National Natural Science Foundation of China
(Grants  No.~11734002, No.~11775021, No.~11874401, No.~11874401, No.~11674406, No.~11822411, No.~12061130200, No.~11227906), and by the
Strategic Priority Research Program (B) of the Chinese Academy of Sciences (Grants No.~XDB25000000, No.~XDB07020000, No.~XDB28000000, 
No.~XDB33010100). H.L. is grateful for support from the Youth Innovation Promotion Association of CAS (Grant No.~2016004) and Beijing Natural Science
Foundation (Grant No.~JQ19002). The work in Boston was supported by the NSF under Grant No.~DMR-1710170 and by the Simons Foundation under
Simons Investigator Award No.~511064. L.L. would like to thank Boston University's Condensed Matter Theory Visitors Program for support. We also acknowledge
the Super Computing Center of Beijing Normal University and Boston University's Research Computing Services for their support.
\end{acknowledgments}

\begin{widetext}

\newpage

\begin{center}  

{\large Supplemental Material}
\vskip3mm

{\bf\large Extreme Suppression of Antiferromagnetic Order and Critical Scaling \\ \vskip1mm
in a Two-Dimensional Quantum Magnet}

\vskip5mm

Wenshan Hong,$^{1,2}$ Lu Liu,$^{1}$ Chang Liu,$^{1,2}$ 2 Xiaoyan Ma,$^{1,2}$ Akihiro Koda,$^{3,4}$ Xin Li,$^{5}$ Jianming Song,$^{5}$
Wenyun Yang,$^{6}$ Peng Cheng,$^{7}$ Hongxia Zhang,$^{7}$ Wei Bao,$^{7,8}$ Xiaobai Ma,$^{9}$ Dongfeng Chen,$^{9}$ Kai Sun,$^{9}$ Wenan Guo,$^{10,11}$
Huiqian Luo,$^{1,12}$ \\ Anders W. Sandvik,$^{13,1,*}$ and Shiliang Li,$^{1,2,12,\dagger}$
\vskip3mm

{\it
{$^1$ Beijing National Laboratory for Condensed Matter Physics,\\ Institute of Physics, Chinese Academy of Sciences, Beijing 100190, China}\\
{$^2$ School of Physical Sciences, University of Chinese Academy of Sciences, Beijing 100190, China} \\
{$^3$ Institute of Materials Structure Science, High Energy Accelerator Research Organization (KEK-IMSS), \\ 1-1 Oho, Tsukuba 305-0801, Japan} \\
{$^4$ Department of Materials Structure Science, Sokendai (The Graduate University for Advanced Studies), \\ Tsukuba, Ibaraki, 305-0801, Japan} \\
{$^5$ Key Laboratory of Neutron Physics and Institute of Nuclear Physics and Chemistry,\\ China Academy of Engineering Physics, Mianyang 621999, China} \\
{$^6$ State Key Laboratory for Mesoscopic Physics, School of Physics, Peking University, Beijing, 100871, China} \\
{$^7$ Department of Physics and Beijing Key Laboratory of Opto-electronic Functional Materials \& Micro-nano Devices, \\
      Renmin University of China, Beijing 1 00872, China} \\
{$^8$ Department of Physics, City Univesity of Hong Kong, Kowloon, Hong Kong}\\
{$^9$ Department of Nuclear Physics, China Institute of Atomic Energy, Beijing, 102413, China} \\
{$^{10}$ Department of Physics, Beijing Normal University, Beijing 100875, China} \\
{$^{11}$ Beijing Computational Science Research Center, Beijing 100193, China} \\
{$^{12}$ Songshan Lake Materials Laboratory, Dongguan, Guangdong 523808, China} \\
{$^{13}$ Department of Physics, Boston University, 590 Commonwealth Avenue, Boston, Massachusetts 02215, USA}\\} 
  
\vskip1mm
$^*$ e-mail: sandvik@bu.edu, $^\dagger$ slli@iphy.ac.cn

\end{center}
\vskip8mm

We provide additional experimental (Sec.~\ref{supp:exp}), theoretical (Sec.~\ref{supp:lambda}), and Monte Carlo simulation
(Sec.~\ref{supp:heisenberg}) results supporting the conclusions of the main paper. Additional $\mu$SR $A(t)$ data are presented in
Sec.~\ref{supp:usr} and the fitting procedures are explained. Neutron diffraction data in the columnar AFM state are presented in Sec.~\ref{supp:neutron}
and in Sec.~\ref{supp:tstar} we explain how the cross-over temperature $T^*$ was determined from the neutron data. In Sec.~\ref{supp:lambda}, we derive the
scaling form of the $\mu$SR relaxation rate $\lambda$. In Sec.~\ref{supp:heisenberg}, we present additional Monte Carlo results for the classical
2D Heisenberg model with W-type impurities.
\vskip10mm

\end{widetext}

\setcounter{page}{1}
\setcounter{equation}{0}
\setcounter{figure}{0}
\renewcommand{\theequation}{S\arabic{equation}}
\renewcommand{\thefigure}{S\arabic{figure}}
\renewcommand{\thesection}{\arabic{section}}

\section{Additional Experimental Information}
\label{supp:exp}

Polycrystalline samples of Sr$_2$CuTe$_{1-x}$W$_x$O$_6$ were synthesized from stoichiometric mixtures of SrO, CuO, TeO$_2$, and WO$_3$ powders by the
solid-state reaction method reported previously \cite{KogaT16,BabkevichP16,VasalaS14,WalkerHC16}. 
The $\mu$SR experiments were performed at the S1 ARTEMIS spectrometer (Proposal No.~2018B0156), J-PARC, with the mini cryostat down to 4 K.
The neutron-diffraction experiments were carried out at Bamboo ($\lambda$ = 2.358 \AA) and Xingzhi  ($\lambda$ = 2.7302 \AA)
triple-axis spectrometers, and at the PKU High-Intensity Powder Neutron Diffractometer ($\lambda$ = 2.3 \AA) at China Advanced Research Reactor (CARR),
and the Kunpeng triple-axis spectrometer ($\lambda$ = 2.7302 \AA) at Key Laboratory of Neutron Physics and Institute of Nuclear Physics and Chemistry, China.
Neutron speed velocity selectors were used before the monochromator with the Bamboo and Xingzhi spectrometers.

\begin{figure*}[t]
\includegraphics[width=123mm]{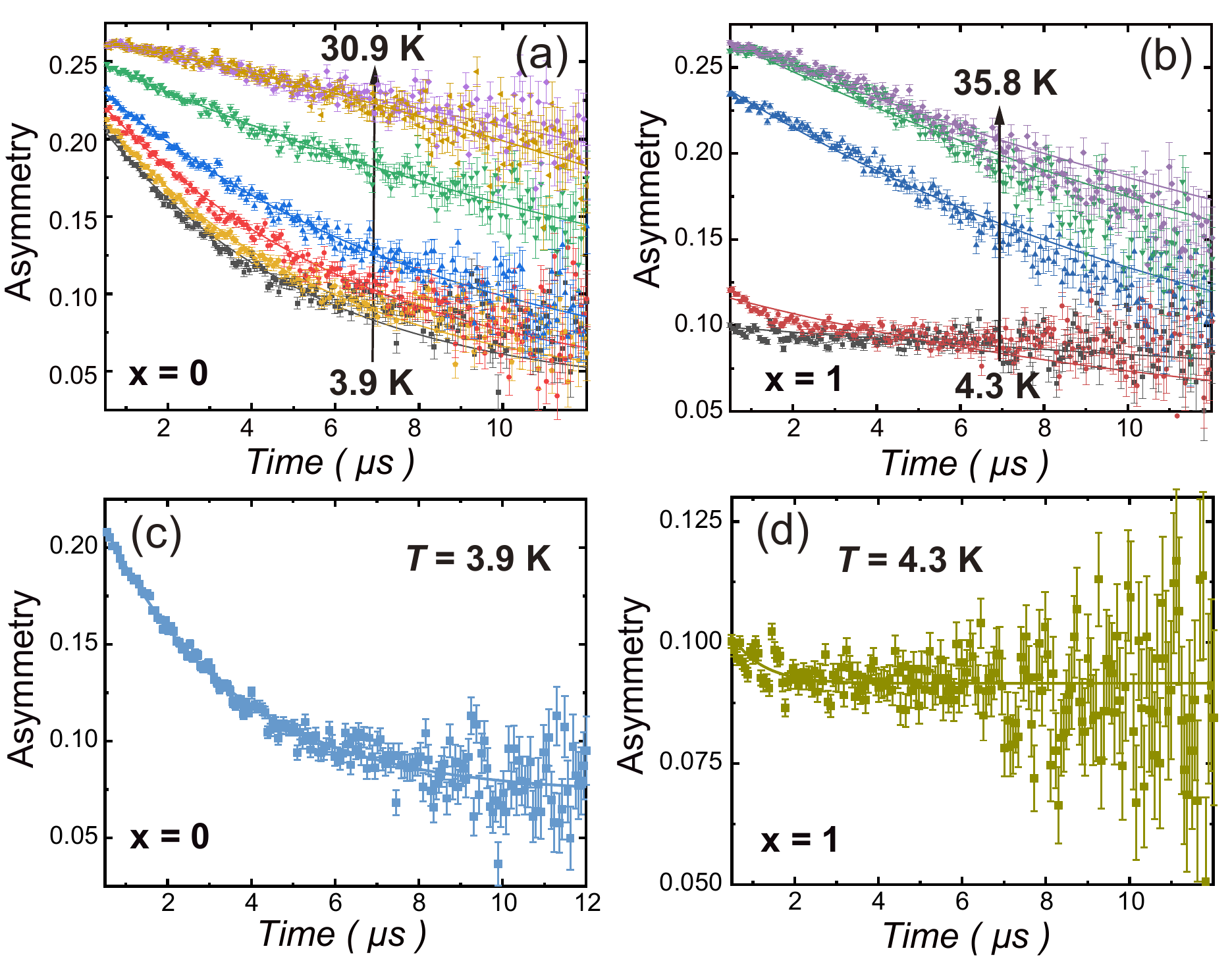}
\caption{(a) and (b) show zero-field $\mu$SR spectra of Sr$_2$CuTe$_{1-x}$W$_x$O$_6$ samples with $x = 0$ and 1, respectively. Results for several temperatures
are shown, witth the highest and lowest indicated for both samples. The curves are fits to the form Eq.~(\ref{atform}) with a single relaxation rate. (c) and (d)
show the spectra for $x = 0$ and 1, respectively, at the corresponding base temperatures. The curves are fits to the modified form Eq.~(\ref{atformnew}), which
provides
a better description of the data in the ordered state.}
\label{rawusrdata}
\end{figure*}

\subsection{Raw $\mu$SR data}
\label{supp:usr}

The time dependent asymmetry $A(t)$ from our $\mu$SR experiments for $x$ = 0 and 1 are shown in Fig.~\ref{rawusrdata}. As discussed in the main text, the
$x = 0$ sample [Fig.~\ref{rawusrdata}(a)] has long-rage N\'eel AFM order, while the $x = 1$ sample [Fig.~\ref{rawusrdata}(b)] has long-range columnar
order. It is clear that the fits by Eq.~(\ref{atform}) are not good at low temperatures. This is in contrast with the good fits at $x$ = 0.05 and 0.1, as shown
in Figs.~\ref{fig:usr}(a) and \ref{fig:usr}(b).

The reason for the suboptimal fits at $x$ = 0 and 1 is that, in the ordered states, we need multiple relaxation rates to describe the data, as shown in
Ref.~\cite{VasalaS14b}. Here we test the following simpler function:
\begin{equation}
A(t) = A_0[f + (1-f){\rm exp}(-\lambda t)]G_z(t) + A_{\rm BG}.
\label{atformnew}
\end{equation}
\noindent Compared to the fitting function in the main text, the new function introduces a factor $f$ to effectively account for a second relaxation rate
that is very small, so that its value is effectively zero on the time scale of the experiment. The very well fitted low-temperature results for $x$ = 0 and 1
are shown in Figs.~\ref{rawusrdata}(c) and ~\ref{rawusrdata}(d). It is worth noting that $f$ is close to 1/3 for $x$ = 0, and 1/2 for $x$ = 1. We stress that we
need the modified fitting form only for analyzing the ordered samples. As noted in the main text and shown in Figs.~\ref{fig:usr}(a) and \ref{fig:usr}(b), for
the short-range correlated samples with $x$ = 0.05 and 0.1 the form Eq.~(\ref{atform}) works essentially perfectly.

When fitting the $\mu$SR spectra, we have chosen a temperature-independent background $A_{\rm BG}$ = 0.035 for all the samples. This value is
derived from the fact that the value of $A(t)$ at 1 $\mu s$ at base temperature is about 1/3 of that above $T_c$, as shown in Ref.~\cite{VasalaS14b}. The same
instrument was used for all the $\mu$SR measurements and all the samples have similar mass and were mounted in similar holders. For all these reasons
we expect that the background should be close to the same for all the samples. Reasonable fits can be obtained for $A_{\rm BG}$ ranging from 0 to 0.1, and
using different values in this range does not affect the conclusion of low-temperature  power-law scaling $\lambda \sim T^{-\gamma}$ for $x$ = 0.05 and 0.1;
the exponent changes only marginally and $\gamma(0.1) > \gamma(0.05)$ always holds.

\begin{figure}[b]
\includegraphics[width=88mm]{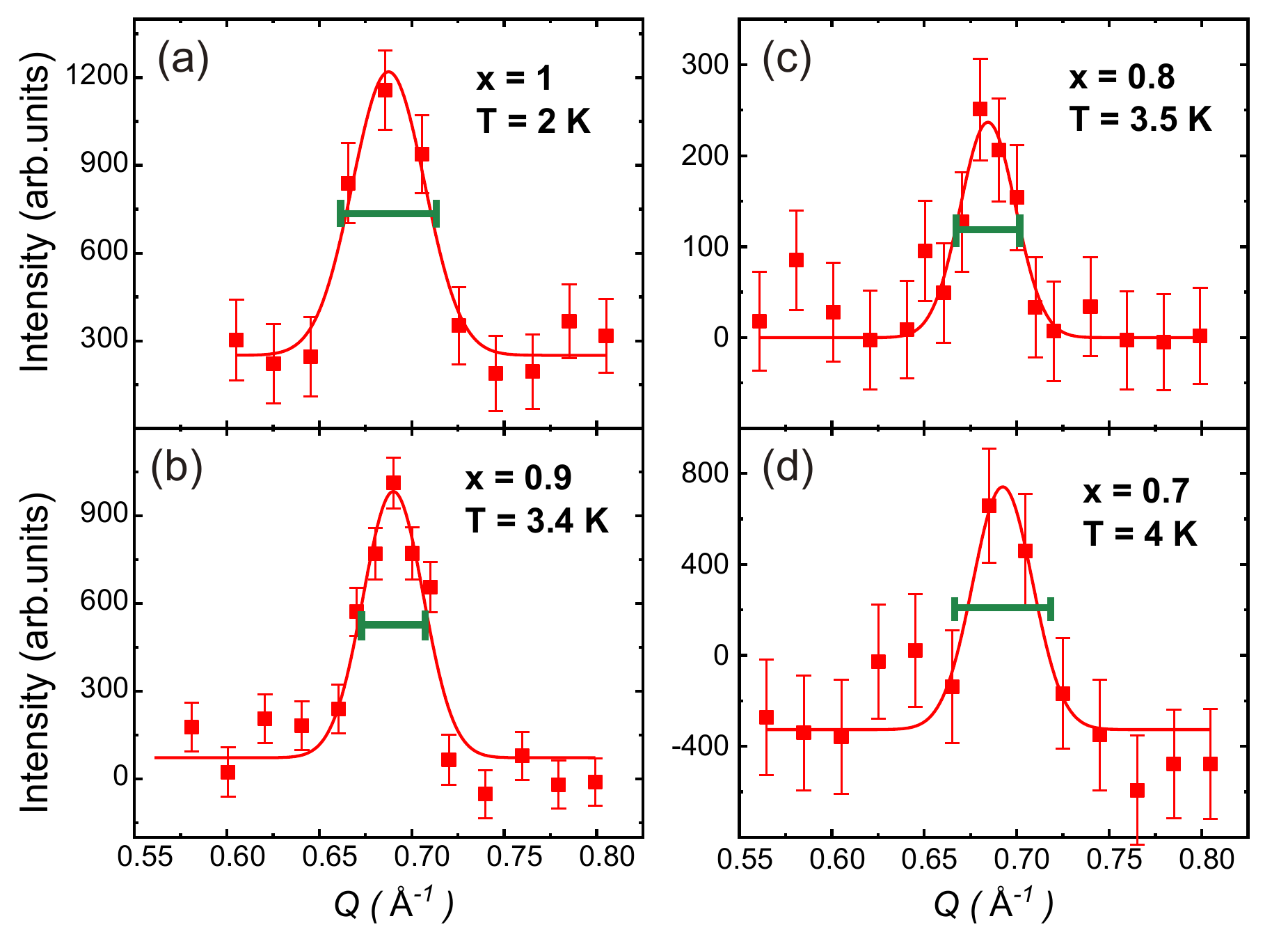}
\caption{Neutron diffraction data along with Gaussian fits for $x=1$ (a) $0.9$ (b), $0.8$ (c), and $0.7$ (d). The peak locations correspond
to $q=(0.5, 0, 0.5)$ and $(0, 0.5, 0.5)$, i.e., columnar AFM structure. The temperature is indicated in each panel and data taken at $T=40$ K have
been subtracted as background contributions. The green bars indicate the instrumental resolution.}
\label{qscans2}
\end{figure}

\subsection{Neutron diffraction results for the columnar AFM state}
\label{supp:neutron}

Neutron diffraction data for $x$ from $0.7$ to $1$ are shown in Fig.~\ref{qscans2}. At these W fractions the system is expected from previous studies
\cite{MustonenO18b} to have columnar AFM order at low temperature, which we confirm here with the resolution limited peaks at the corresponding wave-vectors.

\begin{figure*}
\includegraphics[width=110mm]{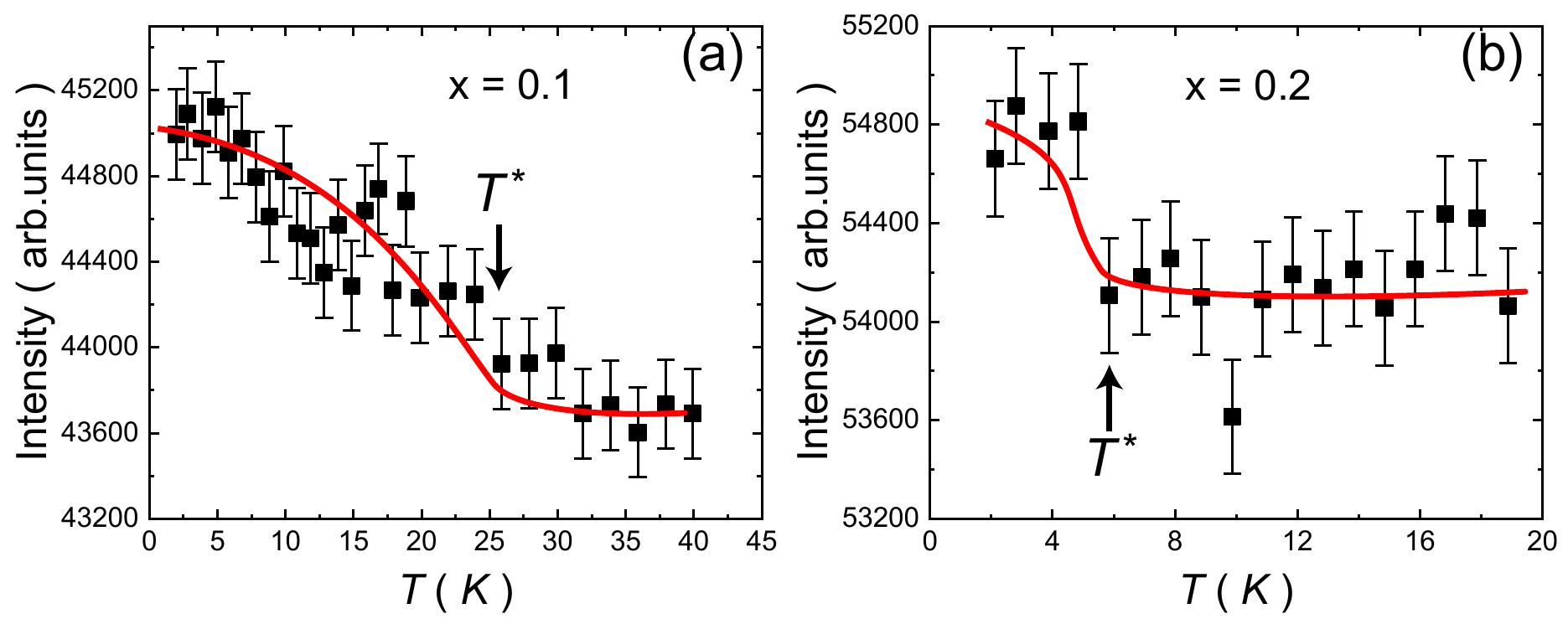}
\caption{Temperature dependence of the magnetic peak intensity at $q =(0.5, 0.5, 0)$ in the sample with $x = 0.1$ in (a) and $x=0.2$ in (b).
The curves are guides to the eye and the cross-over temperature is defined as the point where the signal above background becomes significant,
which implies errors of up to $3$ K in these cases.}
\label{MS}
\end{figure*}

\subsection{The cross-over temperature $T^*$ from neutron diffraction}
\label{supp:tstar}

For the samples showing no phase transition into an ordered phase, in Fig.~\ref{fig:phases} we have indicated
a temperature $T^*$ where both the $\mu$SR and netron data show the onset
of significant short-range correlations. It should be noted that, strictly speaking, $T^*$ can not be defined unambiguously or uniquely as it merely signifies
a sharp cross-over. Therefore, $T^*$ determined from the neutron-diffraction measurements is not necessarily exactly equal to that from the $\mu$SR data,
since these two techniques measure the system in different ways and with very different energy resolution. We here show that both experiments nevertheless
produce compatible results for $T^*$.

Figures \ref{MS}(a) and \ref{MS}(b) show the temperature dependence of the magnetic peak intensity measured with neutron diffraction at wave-vector
$q = (0.5, 0.5, 0)$ (corresponding to N\'eel AFM order) for the $x = 0.1$ and $0.2$ samples, respectively. $T^*$ is determined to be the temperature
where a signal is detected above the high-$T$ background, which is $T^*\approx 25$ and $T^* \approx 6$ K, respectively, for $x=0.1$ and $x=0.2$, with rather
large error bars of $2$-$3$ K due to the weak signal. Comparing with the $\mu$SR result for $x=0.1$ in Fig.~\ref{fig:phases}, the results agree well.
We do not have $\mu$SR results for $x=0.2$.

\section{Critical scaling of the relaxation rate}
\label{supp:lambda}

As discussed in the main paper, the $x=0.05$ and $0.1$ samples exhibit quantum-critical scaling in the $\mu$SR relaxation rate and are candidates for the
RS state at low temperatures. According to QMC simulations of a ``designer model'' relizing the RS phase in a 2D quantum magnet \cite{Liu18,Liu20}, this state
is critical with large dynamic exponent, $z \ge 2$, with $z=2$ at the transition from the N\'eel state and $z$ increasing upon moving into the RS
phase, and with dominant N\'eel type spin correlations decaying with distance $r$ as $r^{-2}$ universally. This correlation function formally implies
that the exponent $\eta$ in the standard form \cite{Fisher89} of the quantum-critical correlation function for a system
in $d$ space dimensions,
\begin{equation}
C(r) \propto r^{-(d+z-2+\eta)},
\label{crform}  
\end{equation}
depends on $z$ through the relationship $\eta=2-z$. Thus, in the RS state this exponent is negative, which is
normally not possible in uniform systems but is not uncommon in disordered systems.

The exponent $\eta$ appears also in dynamical scaling forms,
e.g., the NMR relaxation rate $1/T_1$ scales as $T^\eta$ at the O(3) quantum-critical point in uniform antiferromagnets, where $z=1$ \cite{Chubukov93}. 
One can expect the $\mu$SR relaxation rate $\lambda$, which like $1/T_1$ depends on local low-energy spin fluctuations, to scale in the same way.
However, since the dynamic exponent $z \not=1$ in the RS state, the $T^\eta$ form has to be modified as follows: The correlation length in a quantum-critical
system scales as $\xi \propto T^{-1/z}$, and we can therefore formally express the temperature as $T \propto \xi^{-z}$. For $z=1$, we can write
$\lambda \propto T^\eta \propto \xi^{-\eta}$, and the generalization to $z\not=1$ is obtained by inserting the correct $T$-dependent expression for the
correlation length. Thus, $\lambda \propto \xi^{-\eta} \propto T^{\eta/z}$. Using the form $\eta=2-z$ in the RS state, we expect $\lambda \propto  T^{-\gamma}$,
where we have defined the positive exponent $\gamma={1-2/z}$, with $z \ge 2$. This is the exponent that was extracted from the data fits in Fig.~\ref{fig:usr}(d).

The asymptotic scaling form of $\lambda(T)$ can also be derived in a more transparent way: First, consider the well known NMR spin-lattice relaxation
rate $1/T_1$, which for a spin-isotropic system is given by \cite{Slichter90}
\begin{equation}
\frac{1}{T_1} = \frac{\gamma^2}{2}\sum_{\mathbf{q}} A^2(\mathbf{q}) S(\mathbf{q},\omega_{\rm N}),
\label{t1def}
\end{equation}  
where $\gamma$ is the gyromagnetic ratio, $A_q$ is the Fourier transform of the hyperfine constants describing the coupling between the nuclear and
electronic spins, and $\omega_{\rm N}$ is the field-dependent nuclear resonance frequency. The hyperfine coupling is short-ranged in space, and if the nucleus
considered is in the ion hosting the localized electronic spins (e.g., Cu NMR in the material considered here), it is often sufficient to
consider purely local on-site interactions $A_0$, so that the momentum sum in Eq.~(\ref{t1def}) reduces to $A_0^2 S_0(\omega_{\rm N})$, where
$S_0(\omega)$ is the on-site (single-spin) dynamic structure factor.

Typically, the resonance frequency is much lower than other energy scales in
the system, and the zero-frequency limit can be considered (unless there are significant spin diffusion contributions, which can cause low-frequency
divergencies). Thus, with these simplifications, which are often completely valid, the relaxation rate is proportional to $S_0(\omega \to 0)$
(with prefactors that are known or can be measured). Since $\mu$SR also is a probe of low-frequency local spin fluctuations, we expect the
same form;
\begin{equation}
\lambda \propto S_0(\omega \to 0).
\label{lambdas0}
\end{equation}
  
The local dynamic spin structure factor $S_0(\omega)$ (and also its $q$ dependent variant) can be calculated in various analytical
approximative schemes or numerically; for example, it was calculated in the case of the 1D RS state in Ref.~\onlinecite{Shu18}. However, the
low-frequency limit is often challenging, especially in QMC calculations, where the corresponding imaginary-time dependent spin correlation
function $G_0(\tau)$ has to be calculated and analytically continued to real frequency. To circumvent the latter step, Randeria et al.~suggested
a very useful approximation \cite{Randeria92}, which was expressed in a slightly different form in Ref.~\onlinecite{Shu18}. Neglecting unimportant
factors, the approximation amounts to
\begin{equation}
S_0(\omega \to 0) \propto \frac{1}{T}G_0(\tau=\beta/2),
\end{equation}
and then the relaxation rate Eq.~(\ref{lambdas0}) is approximated as
\begin{equation}
\lambda \propto \frac{1}{T} G_0(\tau=\beta/2),
\label{lambdag0}
\end{equation}
where $\beta=1/T$. Here we will use this form, which is expected in general to become better with decreasing $T$, to derive the critical
scaling behavior of $\lambda$ in the RS phase.

As already mentioned above, a quantum-critical spatial correlation function is conventionally written as Eq.~(\ref{crform}), where $d=2$
in our case. The on-site correlation in imaginary time is modified by the dynamic exponent \cite{Fisher89}
\begin{equation}
G_0(\tau) \propto \tau^{-(d+z-2+\eta)/z},
\label{g0def}
\end{equation}
reflecting that space and (imaginary) time distances are related as $\tau \sim r^z$, which is used to obtain Eq.~(\ref{g0def}) from Eq.~(\ref{crform}).
Thus, in the RS state with the staggered spatial spin correlation
function $C(r) \propto r^{-2}$, the time correlations take the form $G_0(\tau) \propto \tau^{-2/z}$. Using this form in Eq.~(\ref{lambdag0}) immediately
gives the scaling form $\lambda \propto T^{-(1-2/z)}$, in agreement with the result presented earlier. The fact that we observe this kind of scaling
with $z>2$, Fig.~\ref{fig:usr}(d), with $z$ also increasing upon moving further away from the N\'eel phase as predicted \cite{Liu18}, constitutes
strong support for an RS phase in Sr$_2$CuTe$_{1-x}$W$_x$O$_6$.

\section{2D Heisenberg model}
\label{supp:heisenberg}

\begin{figure}[t]
\includegraphics[width=75mm]{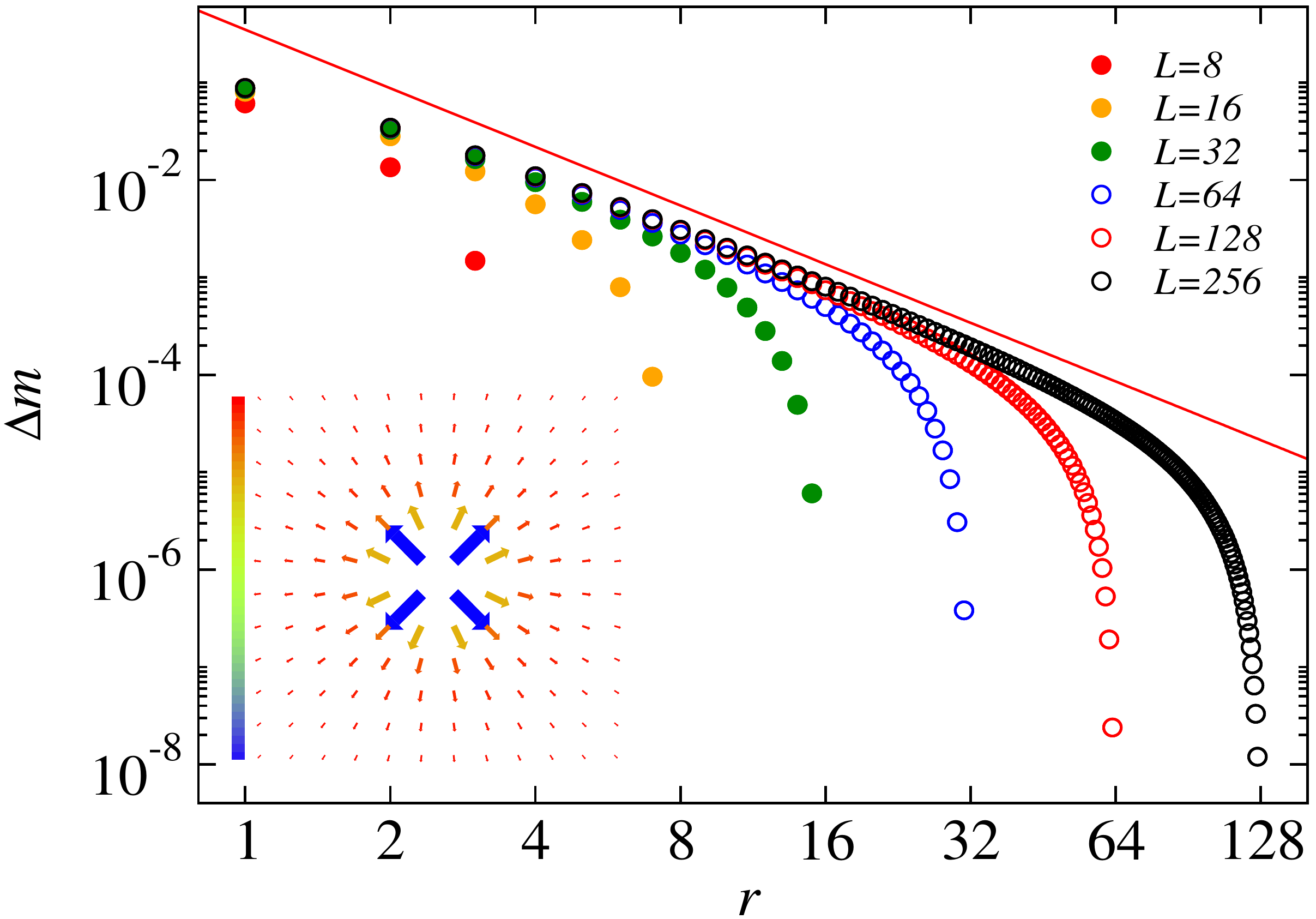}
\caption{Plaquette impurity induced deformation of the N\'eel order parameter for different system sizes, as in Fig.~\ref{fig:deltam} but
with $J'_2=0$. } 
\label{fig:dm0}
\end{figure}

For the Monte Carlo simulations of the classical Heisenberg models, we used methods that have been previously explained in detail in
the literature \cite{Alonso96,Lee07}. The simulations combine heat-bath sweeps with energy conserving ``over-relaxation'' updates. We
found the latter to be particularly important for reaching the ground state of systems with a small number of W-type impurities.
In all simulations, we started at a high temperature and gradually lowered the temperature in order to alleviate problems with long autocorrelation
times. For the systems with more than one W impurity (random mixes of Te and W plaquettes) disorder averages were taken over hundreds of realizations
of random locations of the impurities.

In Fig.~\ref{fig:deltam} in the main text we demonstrated an impurity induced deformation of the sublattice magnetization that decays with the distance
$r$ from the impurity as $1/r^2$. This decay implies that the total response of a single impurity diverges logarithmically with increasing system size.
We here provide additional results demonstrating that the order parameter indeed vanishes for any concentration $x>0$ of the impurities.

In the main paper, the Monte Carlo simulations were carried out with parameters approximating those estimated \cite{Katukuri20} for
Sr$_2$CuTe$_{1-x}$W$_x$O$_6$. The bulk parameters for $x=0$, illustrated in Fig.~\ref{fig:j1j2}(a), were $J_1=1$ and $J'_2=0.1$. Even with the small
frustrating $J'_2$ terms, the $T=0$ order parameter is the fully colinear N\'eel state, and we do not expect that the frustration is in any way required
to obtain the $r^{-2}$ decay of the deformation. To explicitly demonstrate that the classical Heisenberg model with only the first-neighbor
couplings $J_1$ also has the same impurity response as in Fig.~\ref{fig:deltam}, here in Fig.~\ref{fig:dm0} we show simulation results for $J'_2=0$.
These results confirm that the $r^{-2}$ form emerges as the system size increases.

The $1/r^2$ form with no angular dependence of the deformation of the order parameter may appear surprising in light of there being no such
momopole-like solution
of the Poisson equation, which provides the long-distance continuum description of the N\'eel state with impurities \cite{Sushkov03}. As will be discussed
in more detail elsewehere \cite{Liu20b}, the plaquette impurity considered here can be regarded as a composite of two dipoles, with the relative angle of the
deformation vectors in the $xy$ plane chosen to minimize the energy. The angular degree of freedom of the deformation is missing in treatments of impurities
in long-range ordered systems of spins with only two components \cite{Vannimenus89}.

\begin{figure}[t]
\includegraphics[width=75mm]{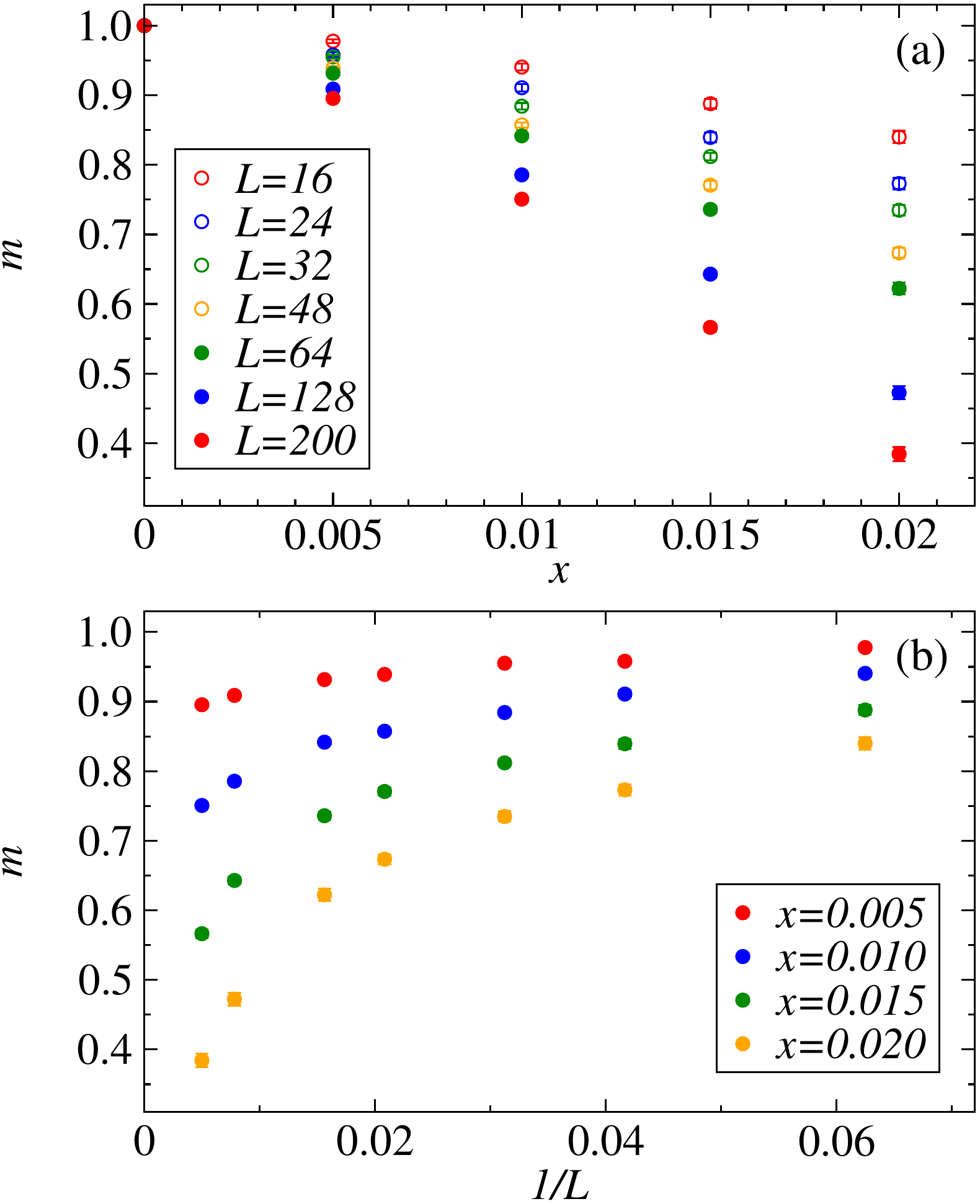}
\caption{(a) Disorder averaged order parameter versus the concentration of W-type plaquette impurities, graphed for several system sizes.
(b) Order parameter at several fixed impurity concentrations $x$ graphed vs the inverse system size.}
\label{fig:magpl}
\end{figure}

For the following results we go back to $J'_2=0.1$, and we expect the same kinds of behaviors also for $J'_2=0$.
In Fig.~\ref{fig:magpl}(a) we show results for the disorder-averaged $T=0$ N\'eel order parameter $m$ versus the concentration of impurities. Increasing
the system size consistently leads to a smaller value of $m$. In Fig.~\ref{fig:magpl}(b) we show results versus the inverse system size for several low
impurity concentrations. Here we can observe that $m$ always decreases with increasing $L$. Given the logarithmic singularity suggested by the single-impurity
response, the most natural scenario is that $m$ vanishes in the thermodynamic limit for all $x >0$, but it is difficult to demonstrate that reliably
using results such as those in Fig.~\ref{fig:magpl}, because of the logarithmic-type singularity that makes extrapolations difficult.

A better way to investigate the presence or absence of order for small $x$, introduced in Ref.~\onlinecite{Liu09}, is to consider a system with a single
impurity to have concentration $x=1/L^2$, and to compute the initial slope,
\begin{equation}
R=\frac{dm}{dx},
\end{equation}
of the order parameter vs $x$ based on this value;
\begin{equation}
R_1(L)=L^2[1-m_1(L)],
\label{r1def}
\end{equation}
where $m_1$ is the value of $m$ computed with the single impurity (averaged over the entire system). Then, if indeed $m=0$ for $L \to \infty$ at $x=0^+$,
the slope $R_1(L)$ will diverge. In order to take into account possible subtle interaction effects, we here additionally use a modified approach
with $L$ randomly placed impurities in the $L^2$ system, for which the concentration is $x=1/L$ and the slope is
\begin{equation}
R_L(L)=L[1-m_L(L)],
\label{rldef}
\end{equation}
where $m_L(L)$ is the impurity-averaged order parameter for $L$ impurities in the lattice with $L^2$ spins.

\begin{figure}[t]
\includegraphics[width=75mm]{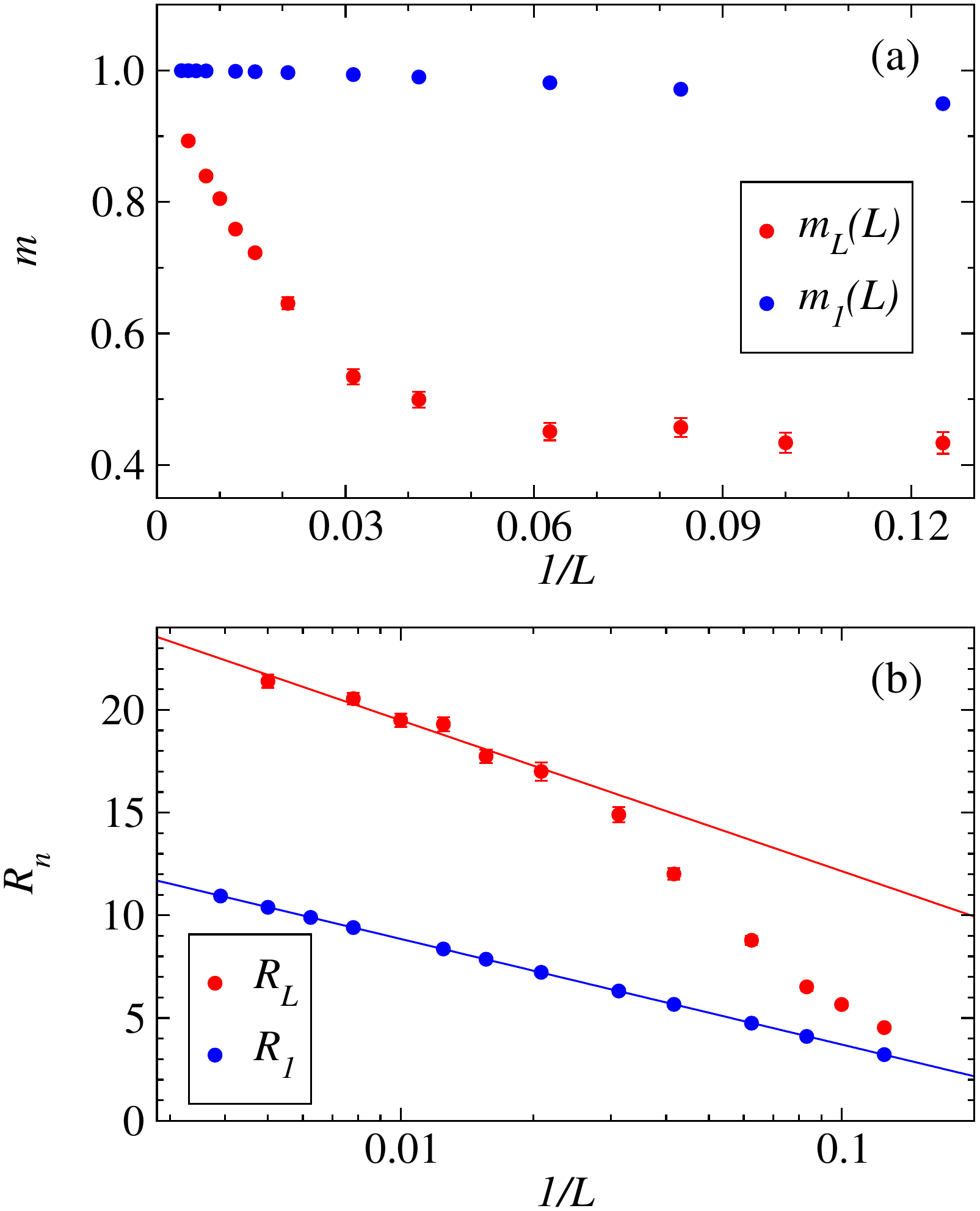}
\caption{(a) N\'eel order parameter vs inverse system size in systems with a single impurity (blue symbols) and with $L$ impurities (red symbols), graphed versus
the inverse of the system size $L$. (b) Slope graphed on a log-linear plot of the magnetization curve at $x=0$ based on the size-dependent definitions, Eqs.~(\ref{r1def})
and (\ref{rldef}), with the data in (a). The lines are fits corresponding to the logarithmically divergent forms $R_n(L) \sim a_n + b_n\log(L)$ with both definitions
(with systems containing $n=1$ and $n=L$ impurities).}
\label{fig:m1ml}
\end{figure}

In Fig.~\ref{fig:m1ml}(a) we show $m_1(L)$ and $m_L(L)$ versus $1/L$. In the former, we can see clearly the expected approach to the fully saturated
bulk order parameter $m=1$ when $L$ increases. For $m_L(L)$ we also have to asymptotically approach the same limit, and this appears plausible though the convergence is
slower, as expected, because of the higher concentration $x$ for a given system size.
In Fig.~\ref{fig:m1ml}(a) we graph the initial slopes defined in Eqs.~(\ref{r1def}) and (\ref{rldef}). Both quantities diverge
logarithmically, confirming that the impurity response in the $x\to 0$ limit has a logarithmic singularity. Any other interpretation than
$m(x)=0$ for all $x>0$ is then unlikely, as indicated also by the results in Fig.~\ref{fig:m1ml} for small but finite impurity concentrations.

The N\'eel order suppression for any $x>0$ is also supported by the strong sensitivity of $T_c(x)$ to the 3D coupling $J_\perp$ in Fig.~\ref{fig:phases}(b), which
suggests that the transition into the ordered phase at $x>0$ and $T>0$ is due to the inter-layer effect. It would be intersting to also study the deformation
induced by a single-impurity in the 3D coupled-layer system, but we have not yet done so. We should expect the $1/r^2$ decay to be cut off at some distance
depending on $J_{\perp}$ (diverging as $J_{\perp}/J_1 \to 0$) and, therefore, the slopes defined in Eqs.~(\ref{r1def}) and (\ref{rldef}) to be finite for any
$J_\perp > 0$.

Related issues were recently discussed by Dey et al.~in the context of a host system (the Heisenberg model on the triangular lattice) with coplanar AFM
order \cite{Dey20}. While previous works have considered distruction of long-range order by dipolar impurities in two-component spin systems (the XY model)
\cite{Vannimenus89}, this system lacks the rotational degree of freedom of the distortion field of impurities in the Heisenberg case. The lack of previous
works on the plaquette impurity (which, as we pointed out, can be regarded as a composite of two dipoles  at a certain relative angle) likely reflects
the absence of experimental motivation before the investigations of Sr$_2$CuTe$_{1-x}$W$_x$O$_6$ demonstrated these particular coupling patterns
\cite{MustonenO18a,WatanabeM18,MustonenO18b,Katukuri20}.

\vfill
\end{document}